\documentclass[conference,preprint]{IEEEtran}
\IEEEoverridecommandlockouts

\usepackage[utf8]{inputenc}
\usepackage[pdftex]{graphicx}

\usepackage{textcomp}

\usepackage{amsmath,amssymb,amsfonts,url,multirow}
\usepackage{algorithmicx}
\usepackage{algpseudocode}
\usepackage{cases}
\usepackage{ulem}
\usepackage{epstopdf}
\usepackage{listings}
\usepackage{array}
\usepackage{makecell}
\usepackage{tabularx}
\usepackage{multicol}

\usepackage[x11names,table]{xcolor}
\usepackage{colortbl}

\usepackage{caption}
\usepackage{booktabs}

\include{timeline}

\usepackage[edges]{forest}

\usepackage{pifont}

\def\BibTeX{{\rm B\kern-.05em{\sc i\kern-.025em b}\kern-.08em
    T\kern-.1667em\lower.7ex\hbox{E}\kern-.125emX}}

\newcommand\RSSI{\textrm{RSSI}}
\newcommand\PRR{\textrm{PRR}}

\newcommand\RSSIavg{\textrm{RSSI}_\textrm{avg}}
\newcommand\RSSIstd{\textrm{RSSI}_\textrm{std}}

\newcommand\Wprr{\textrm{W}_\textrm{PRR}}
\newcommand\Whistory{\textrm{W}_\textrm{history}}

\ifCLASSOPTIONcompsoc
	\usepackage[nocompress]{cite}
	\usepackage[caption=false,font=normalsize,labelfont=sf,textfont=sf]{subfig}
\else
	\usepackage{cite}
	\usepackage[caption=false,font=footnotesize]{subfig}
\fi

\definecolor{awesome}{rgb}{1.0, 0.6, 0.0}

\makeatletter
\def\ps@IEEEtitlepagestyle{%
	\def\@oddfoot{\mycopyrightnotice}%
	\def\@evenfoot{}%
}
\def\mycopyrightnotice{%
	{\begin{minipage}{2\linewidth}\footnotesize\bfseries \copyright 2020 IEEE.  Personal use of this material is permitted.  Permission from IEEE must be obtained for all other uses, in any current or future media, including reprinting/republishing this material for advertising or promotional purposes, creating new collective works, for resale or redistribution to servers or lists, or reuse of any copyrighted component of this work in other works.\hfill\end{minipage}}
	\gdef\mycopyrightnotice{}
}

\begin{document}

\title{On Designing a Machine Learning Based Wireless Link Quality Classifier}

\author{Gregor Cerar$^{\ast\dagger}$,
     	Halil Yetgin$^{\ast\ddagger}$,
        Mihael Mohor\v ci\v c$^{\ast\dagger}$,
        Carolina Fortuna$^{\ast}$\\
$^{\ast}$Department of Communication Systems, Jo{\v z}ef Stefan Institute, SI-1000 Ljubljana, Slovenia.\\
$^{\dagger}$Jo{\v z}ef Stefan International Postgraduate School, Jamova 39, SI-1000 Ljubljana, Slovenia.\\
$^{\ddagger}$Department of Electrical and Electronics Engineering, Bitlis Eren University, 13000 Bitlis, Turkey.\\
\{gregor.cerar $\mid$ halil.yetgin $\mid$ miha.mohorcic $\mid$ carolina.fortuna\}@ijs.si    
}

\maketitle

\begin{abstract}
	Ensuring a reliable communication in wireless networks strictly depends on the effective estimation of the link quality, which is particularly challenging when propagation environment for radio signals significantly varies. In such environments, intelligent algorithms that can provide robust, resilient and adaptive links are being investigated to complement traditional algorithms in maintaining a reliable communication. In this respect, the data-driven link quality estimation (LQE) using machine learning (ML) algorithms is one of the most promising approaches. In this paper, we provide a quantitative evaluation of design decisions taken at each step involved in developing a ML based wireless LQE on a selected, publicly available dataset. Our study shows that, re-sampling to achieve training class balance and feature engineering have a larger impact on the final performance of the LQE than the selection of the ML method on the selected data. 
\end{abstract}

\IEEEpeerreviewmaketitle


\section{Introduction}
\label{sec:intro}
Machine learning (ML) is becoming an increasingly popular way of solving various aspects in communications in general and wireless networks in particular. Data driven link quality estimation (LQE) techniques where the researchers manually developed models have been proposed over the last two decades~\cite{nguyen1996trace, balakrishnan1998explicit, woo2003taming}. More recently, the manual model development is being automated, by using ML algorithms that approximate the distribution of the underlying random variable and are thus able to learn the quality of a link~\cite{sun2017wnn,demetri2019automated}. 

LQE developed using ML can estimate the quality of a link in a continuous value space, in this case the ML performs a regression~\cite{liu2011foresee, millan2015time, sun2017wnn, ancillotti2017reinforcement, okamoto2017machine, bote2018online}. Alternatively, if they estimate the quality in a discrete value space, the ML performs classification~\cite{liu2014temporal, rekik2015fli,luo2019link, demetri2019automated}. By analyzing the existing body of work developing classification models for LQE, we notice the following approaches: binary, two class or multi-class. 

The first type is a \textit{binary or a two-class output}, which is produced by the classification model. This type of output can be found in~\cite{woo2003taming, baccour2011radiale, guo2013fuzzy, liu2014temporal, rekik2015fli}. The applications noticed are mainly (binary) decision making~\cite{woo2003taming} and above/below threshold estimation~\cite{baccour2011radiale, guo2013fuzzy, liu2014temporal, rekik2015fli}.

The second type is \textit{multi-class output} value. Similar to the first type, it is also produced by the classification model. The multi-class output values are utilized in~\cite{boano2010triangle, rehan2016machine, shu2017research, audeoud2018quick, luo2019link, demetri2019automated}, where \cite{rehan2016machine,audeoud2018quick} use a three-class, \cite{boano2010triangle} utilizes a four-class, \cite{shu2017research, luo2019link} rely on a five-class, and \cite{demetri2019automated} leverages a seven-class output. The applications observed are the categorization and estimation of the future LQE state, which is expressed through labels/classes. It is not always clear from the related work how the authors select the number of classes. However, according to~\cite{baccour2012radio}, a wireless link seems to follow a non-linear S-shaped curve with three regions. All works using a three class output model seem to consider this characteristic of the link. 

For developing ML models in any application area, generally some very precise steps that are well established in the community are followed~\cite{fayyad1996data,kulin2016data}, namely data pre-processing, model building and model evaluation. The data preprocessing stage is known to be the most time-consuming process and tends to have a major influence on the final performance of the model. This stage includes several steps such as data cleaning and interpolation, feature selection and re-sampling. While most of the identified research developing LQEs explicitly mention aspects of cleaning and interpolation and feature selection, none evaluate the impact of the design decision taken at these steps on the final performance of the ML based LQE. However, the majority evaluate the impact of the ML method selection on the final performance of the ML based LQE.

Additionally, none of the works mentions aspects of the re-sampling step, that is particularly critical for the generalization capability of a model. Re-sampling is used in ML communities when the available input data is imbalanced \cite{chawla2004special, yap2014application}. For instance, assume a classification problem where the aim is to classify links into \textit{good}, \textit{bad} and \textit{intermediate} classes, similar to the problem approached in \cite{rehan2016machine,audeoud2018quick}. If the \textit{good} class would represent 75\% of the examples in the training dataset, \textit{bad} would represent 20\% and \textit{intermediate} would represent the remaining 5\%, then a ML model would likely be well trained to recognize the \textit{good} as it has been exposed to many such instances, however it might have difficulties in recognizing the other two minority classes.

In this paper, we aim to show the impact of design decisions taken at each step of the process of designing a ML based LQE model on the final performance of the model. To realize our aim, we first select the Rutgers publicly available dataset~\cite{kaul2006creating} and a decision tree as a representative ML based classification model. Then, we systematically perform each step of the knowledge discovery process~\cite{fayyad1996data} on the selected dataset using the selected model, meanwhile varying the design parameters at each step. This way, we are able to systematically quantify the influence of each of the design steps on the final performance of our classifier, therefore providing an in-depth step by step understanding on the process of learning to classify wireless links. 

The main contributions of this paper are:
\begin{itemize}
	\item A systematic quantification of the influence of the design steps on the final performance of our wireless link quality classifier is provided. The highlights of the quantification are that, for the chosen problem and dataset, the generation of synthetic features from the only available training feature $RSSI$, yields up to 6\% higher accuracy and is able to better discriminate the intermediate class up to 49\%. The choice of ML method has less, relatively smaller impact on the final model performance with all the selected algorithms yielding an accuracy performance between 94\% and 95\% and minority class is detected between 87\% and 89\%.
	\item A first time evaluation of the impact of re-sampling on wireless link quality classification is realized using ML. In the case of the chosen imbalanced dataset, by using standard re-sampling, the minority class is correctly detected in over 87\% of the instances, yielding more than 25 percentage points increase in the performance and comes at a small 2\% decrease in overall accuracy.
\end{itemize}

The remainder of the paper is organized as follows. Section \ref{sec:rutgers} elaborates on the selected dataset, Section \ref{sub:cleaning-interpolating} analyses the importance of the cleaning and interpolation, Section \ref{sub:feature-selection}, analyses the importance of feature engineering and its sub-steps, while Section \ref{sub:building-model} analyses the importance of the model selection on the final performance of the LQE classifier. Finally, Section \ref{sec:conclusion} concludes the paper.

\section{Rutgers dataset summary}
\label{sec:rutgers}
The Rutgers trace-set \cite{kaul2006creating} includes 4,060 distinct link traces, which are gleaned from 812 unique links with 5 different noise levels, i.e., 0, -5, -10, -15 and -20 dBm. Readily available trace-set features include raw RSSI, sequence numbers, source node ID, destination node ID and artificial noise levels. In this particular experiment, we observe that the packets are sent every 100 milliseconds for a period of 30 seconds. Therefore, each trace is composed of 300 packets. Besides, based on the specifications of the radio used, each RSSI value is defined between 0 and 128, where the value of 128 indicates an error and is therefore invalid. Nonetheless, a statistical analysis of the Rutgers trace-set reveals that 960 link traces out of 4,060 (23.65\%) are entirely empty indicating no packets were received, and that a total of 1,218,000 packets were sent and only 773,568 (63.51\%) were correctly received.

All the scripts developed for the comparative performance analyses are publicly available on the GitHub repository\footnote{https://github.com/sensorlab/link-quality-estimation} for researchers to reproduce, re-use on other data-sets and improve upon our analyses.

\section{Analysis of cleaning \& interpolation steps}
\label{sub:cleaning-interpolating}

The first step of the process of developing a ML based LQE involves data cleaning and interpolation. The reason for that is because models that are automatically created using ML algorithms can be significantly biased as a result of invalid and missing data. First of all, a valid time series corresponding to each link has to be extracted, which is referred to as a series of ordered tuples each of which contains a packet sequence number and corresponding measured link metrics. The obtained values in the tuples have to be within valid ranges. For instance, the sequence numbers have to be identical with the packets sent during the trace collection, and the values of the link metrics have to remain within the valid ranges that are specified by the transceiver data sheets. Roughly speaking, link metrics with regard to the received radio signals, i.e., RSSI and LQI can be extracted directly from the hardware registers of the corresponding transceivers, whereas link metrics concerning packet data transmission, i.e., PRR and PSR are computed with suitable software procedures.

As described in Section~\ref{sec:rutgers}, the Rutgers trace-set contains invalid values and a considerable number of missing sequence numbers due to the lost packets. Most of the available out-of-the-box data mining algorithms cannot handle these invalid values, e.g., \textit{NaN} and $\pm\infty$ of IEEE~754 standard, or they are simply ignored. To quantify the impact of selected cleaning and interpolation approaches to the final performance of the model, we assume the use of a decision tree algorithm trained with a trio of instant RSSI, averaged RSSI and standard deviation RSSI values, stratified k-fold and pruning, standard normalization, and random oversampling approach, as discussed in Sections~\ref{sub:feature-selection},~\ref{sub:window-selection} and~\ref{sub:resampling}.

From the perspective of data preprocessing steps for ML models, there are many approaches for handling missing data~\cite{jerez2010missing, little2014statistical}. To reveal the impact of the approach to missing values on link quality classification, we train the same model, i.e., decision trees, stratified k-fold and pruning, along with the same feature set for the following cases; a) without handling the missing values, b) using a simple time series approach where we interpolate missing data with Gaussian noise, and c) with the aid of domain knowledge. In the case of interpolation with Gaussian noise, gaps of missing data are filled with random values based on the previous and next valid values. Regarding domain knowledge, we replace the missing RSSI values with 0, which represents a poor quality link with no received signal, yielding PRR equal to 0. Recalling that possible RSSI values are integers ranging between 0 (bad link with no signal) and 127 (good link with strong signal), while observed value of 128 represents an error. 

\begin{figure}[!htb]
	\subfloat[No interpolation]{\includegraphics[width=0.33\columnwidth]{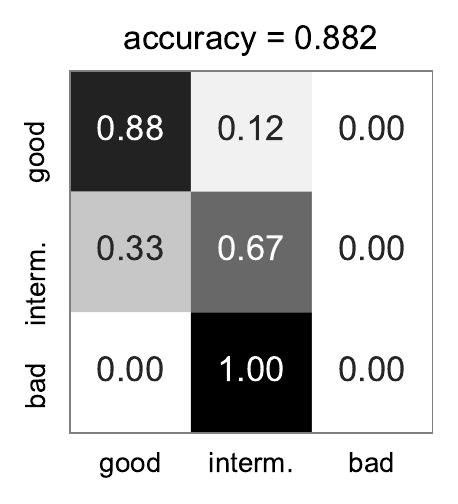}\label{fig:interp:none}}%
	\subfloat[Gaussian]{\includegraphics[width=0.33\columnwidth]{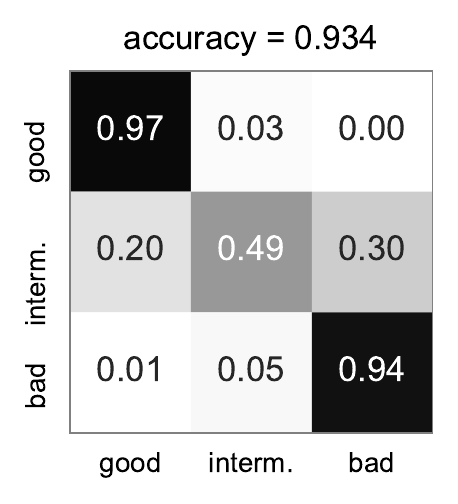}\label{fig:interp:gaussian}}%
	\subfloat[Domain knowledge]{\includegraphics[width=0.33\columnwidth]{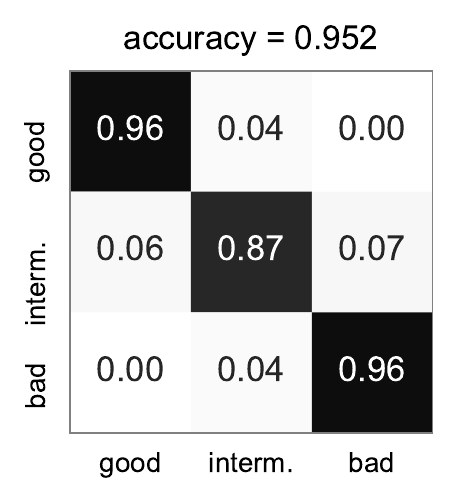}\label{fig:interp:constant}}%
	\caption{Different interpolation cases with a nonlinear decision tree algorithm and random oversampling.}
	\label{fig:interpolations}
\end{figure}

Fig.~\ref{fig:interpolations} presents the relative performance of the models for all three interpolation cases using the form of a confusion matrix\footnote{Confusion matrix is a table layout, with rows for the instances of a predicted class and columns for the instances of an actual class, used for the problem of statistical classification in order to exhibit the performance of an algorithm. Readers are referred to~\cite{Castro8657936} and~\cite{Alshinina8373695} for further details.}, i.e., indicating how well the model classifies individual instances. The better the classifier the darker the diagonal of the confusion matrix and the whiter the non-diagonal squares. For this particular case, the best performing model in terms of accuracy (95.2\%) is the model using domain knowledge in Fig.~\ref{fig:interp:constant}. \textit{While the difference in accuracy between the best two models is approximately 2 percentage points, their respective confusion matrices indicate that the model using interpolation with domain knowledge is superior as it better discriminates between the three link types.} This comparison confirms that \cite{liu2012talent, liu2014temporal,luo2019link} took the best design decision by using domain knowledge for cleaning and interpolation. 

\section{Analysis of feature engineering}
The second step of the process of developing a ML based LQE involves feature engineering. The feature engineering step may involve several sub-steps depending on application requirements, type of data and type of ML problem. For our purpose of learning to classify LQE, we distinguish three  sub-steps discussed in the following subsection.
\subsection{Analysis of feature selection}
\label{sub:feature-selection}
\begin{figure*}[!htb]
	\centering
	\subfloat[$\RSSI$]{\includegraphics[width=0.15\linewidth]{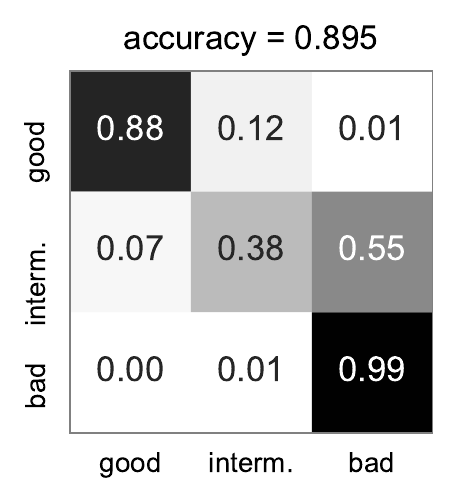}\label{fig:features:dtree:rssi}}%
	\subfloat[$\nabla\RSSI$]{\includegraphics[width=0.15\linewidth]{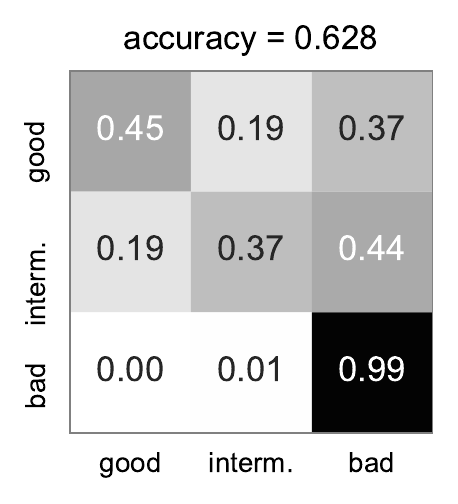}\label{fig:features:dtree:drssi}}%
	\subfloat[$\RSSI^2$]{\includegraphics[width=0.15\linewidth]{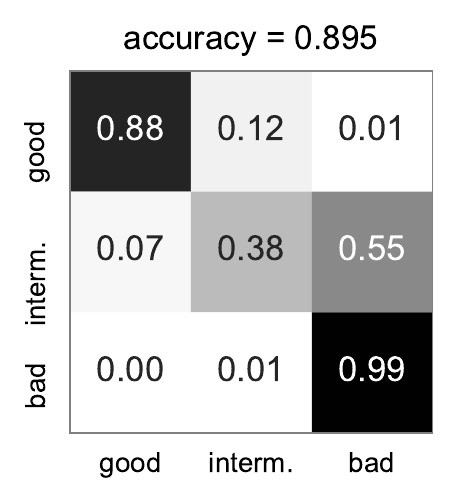}\label{fig:features:dtree:rssi^2}}%
	\subfloat[$\RSSI^3$]{\includegraphics[width=0.15\linewidth]{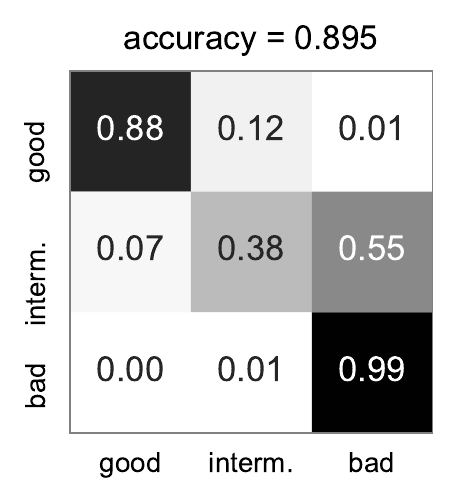}\label{fig:features:dtree:rssi^3}}%
	\subfloat[$\RSSI^4$]{\includegraphics[width=0.15\linewidth]{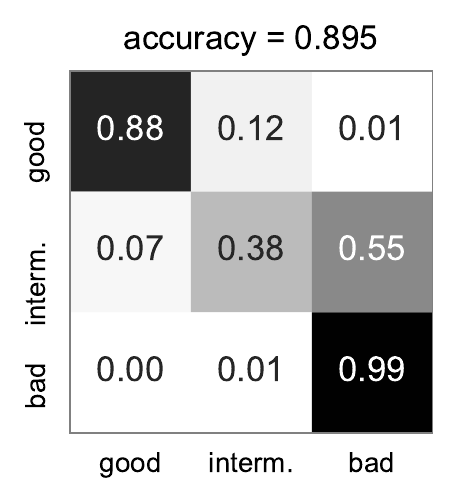}\label{fig:features:dtree:rssi^4}}%
	
	\subfloat[$\RSSI^{-1}$]{\includegraphics[width=0.15\linewidth]{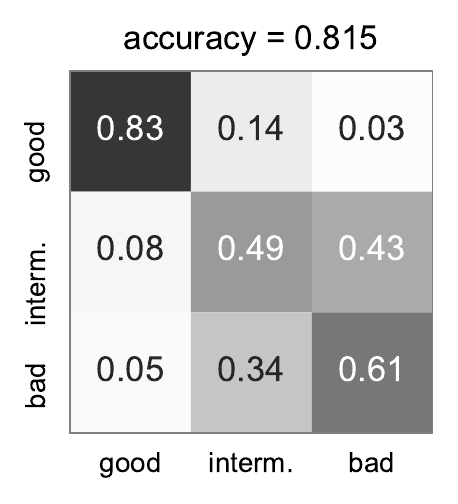}\label{fig:features:dtree:rssi^-1}}%
	\subfloat[$\RSSI^{-2}$]{\includegraphics[width=0.15\linewidth]{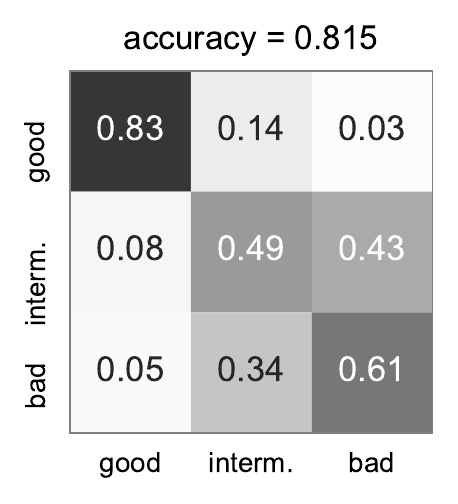}\label{fig:features:dtree:rssi^-2}}%
	\subfloat[$\RSSI^{-3}$]{\includegraphics[width=0.15\linewidth]{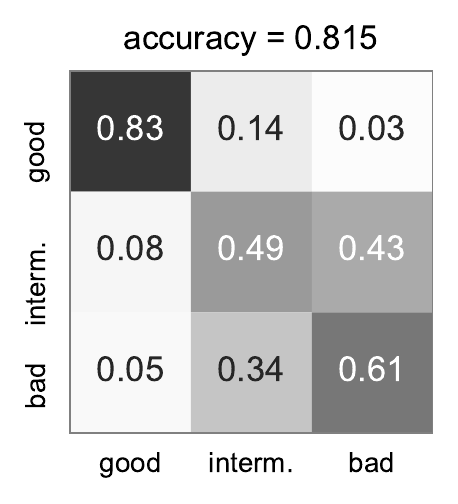}\label{fig:features:dtree:rssi^-3}}%
	\subfloat[$\RSSI^{-4}$]{\includegraphics[width=0.15\linewidth]{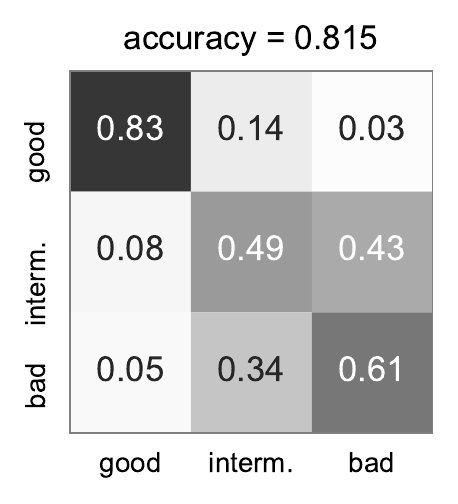}\label{fig:features:dtree:rssi^-4}}%
	\subfloat[$\RSSIavg$]{\includegraphics[width=0.15\linewidth]{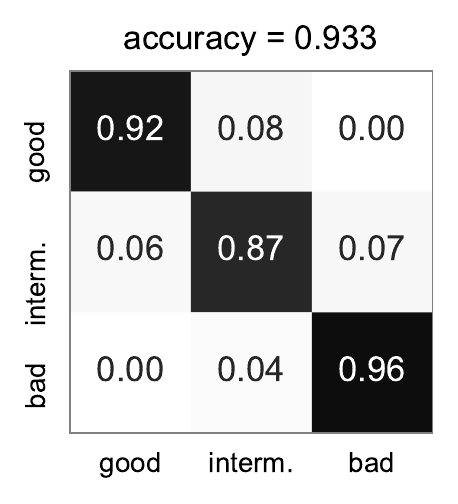}\label{fig:features:dtree:rssi_avg}}%
	
	\subfloat[$\RSSI,\RSSIavg$]{\includegraphics[width=0.15\linewidth]{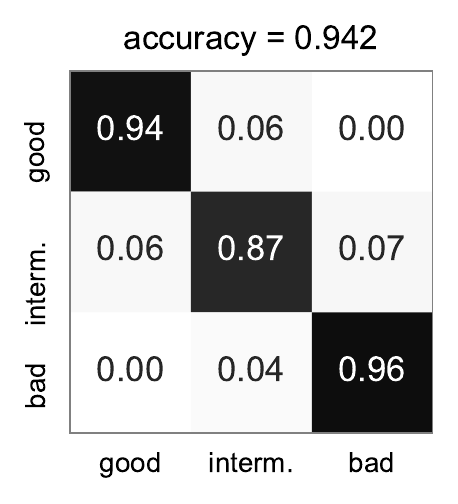}\label{fig:features:dtree:rssi_rssi_avg}}%
	\subfloat[$\RSSI,\RSSIavg,\RSSIstd$]{\includegraphics[width=0.15\linewidth]{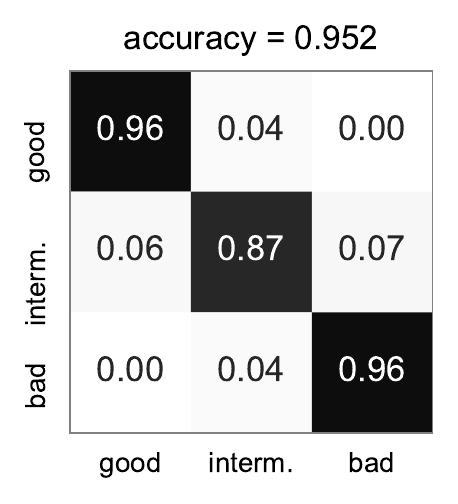}\label{fig:features:dtree:rssi_rssi_avg_std}}%
	\subfloat[$\RSSIavg^{\{1,2,3,4\}}$]{\includegraphics[width=0.15\linewidth]{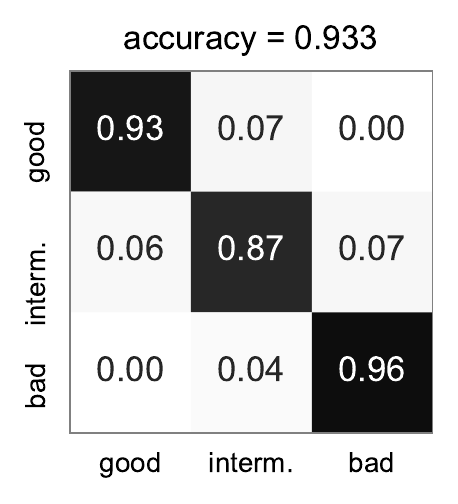}\label{fig:features:dtree:rssi_avg^1_2_3_4}}%
	\subfloat[$\RSSIavg^{\{1,-1,-2,-3,-4\}}$]{\includegraphics[width=0.15\linewidth]{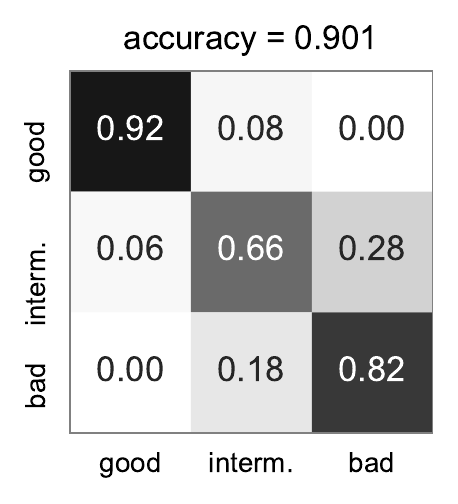}\label{fig:features:dtree:rssi_avg^1_-1_-2_-3_-4}}%
	\subfloat[$\RSSIavg^{\{-4,-3,-2,-1,1,2,3,4\}}$]{\includegraphics[width=0.15\linewidth]{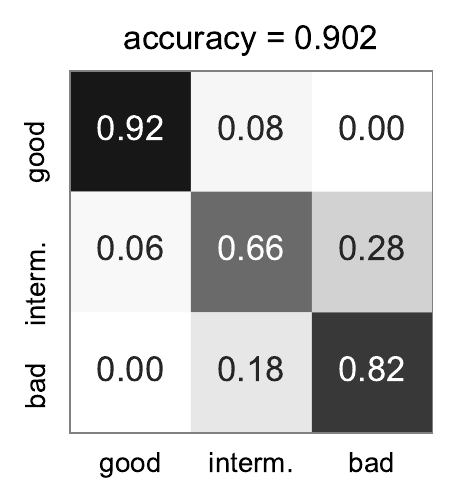}\label{fig:features:dtree:rssi_avg^all}}%

	\caption{The influence of feature selection on the performance of the nonlinear model using decision trees.}
	\label{fig:features-dtree}
\end{figure*}

Feature selection is the process of selecting relevant raw features and/or creating synthetic features to be used for training ML models. When the number of possible input features is very large, then usually only the most relevant ones are selected to be used for the model. On the other hand, when the number of possible input features is very low, then creating synthetic features starting from the available ones to aid in better model development is employed. Feature selection is a fundamental step and can be performed manually or, in some cases, can be built automatically by existing algorithms such as SVMs. 

To analyze and understand the influence of feature selection on the model performance, we consider a set of standard feature engineering procedures on the selected dataset. The Rutgers trace-set has only two available attributes useful for LQE, i.e., the instant (raw) RSSI value and the sequence number, therefore exploring synthetic feature generation for model improvement seems to be the only feasible option for this step. The sequence number is leveraged for the computation of PRR which represents the target value, therefore leaving only RSSI as possible training feature. This classification obeys the following rules:

\begin{equation}\label{eq:prr-to-class}
y = f(\PRR) = \begin{cases}
\text{bad}, & \text{if } \PRR \leq 0.1 \\
\text{intermediate}, & \text{otherwise} \\
\text{good}, & \text{if } \PRR \geq 0.9,
\end{cases}
\end{equation}

\begin{equation}\label{eq:target-class}
\mathbf{y} = [y_1, y_2, \dots, y_n],\quad\forall{y}\in\{\text{bad},\hspace{0.2mm} \text{intermediate},\hspace{0.2mm} \text{good}\}.
\end{equation}

A typical approach in ML for such limited trace-sets is to investigate whether synthetic features, such as average RSSI over a time window or polynomial interactions~\cite{freitas2001understanding}, can aid in training to acquire more accurate models compared to that of the instant RSSI values. Fig.~\ref{fig:features-dtree} shows the influence of the best-performing feature combinations on the classification performance. For this analysis, we assume interpolation based on domain knowledge, i.e., replacing missing values with zeros, as discussed in Section~\ref{sub:cleaning-interpolating}. Additionally, synthetic feature creation with the window sizes $\Wprr$ and $\Whistory$ are set to 10, while utilizing standard normalization and random oversampling approach, as discussed in Sections~\ref{sub:resampling} and~\ref{sub:window-selection}. In this analysis we predict the link quality as per Eq.~(\ref{eq:prr-to-class}) for the next prediction window $\Wprr$.

We can see from the results listed in Fig.~\ref{fig:features-dtree}(a) that the decision tree based model, trained using stratified k-fold and pruning, that uses the only available feature, $RSSI$, yields 89\% accuracy and 38\% correctly identified \textit{intermediate} class, we can call this the baseline performance. \textit{The best performing feature combination that uses two synthetically generated features in $\RSSIavg$, $\RSSIstd$ addition to $\RSSI$ yields an accuracy of 95.2\% and 87\% correctly identified \textit{intermediate} class as can be seen in Fig.~\ref{fig:features-dtree}(l).} Moreover, Fig.~\ref{fig:features-dtree}(j) also shows that $\RSSIavg$ alone yields great results, i.e., 93\% accuracy and 87\% correctly identified \textit{intermediate} class. Positive powers of $RSSI$ have no major advantage over the baseline as can be seen from Fig.~\ref{fig:features-dtree}(c), (d) and (e), while negative powers lower overall accuracy, albeit they perform better then the baseline for the \textit{intermediate} class and significantly worse for the \textit{bad} class as per Fig.~\ref{fig:features-dtree}(f), (g), (h) and (i). In the last row of the table, Figs.~\ref{fig:features-dtree}(k-o) show that other synthetic combinations of $\RSSIavg$ perform relatively better than the baseline. \textit{As a conclusion, it can be seen that the generation of synthetic features from the only available training feature $RSSI$, yields up to 6\% higher accuracy and is able to discriminate the intermediate class up to 49\% better.}

\subsection{Analysis of window selection}
\label{sub:window-selection}

\begin{figure*}[!htb]
	\centering
	\includegraphics[width=0.7\linewidth]{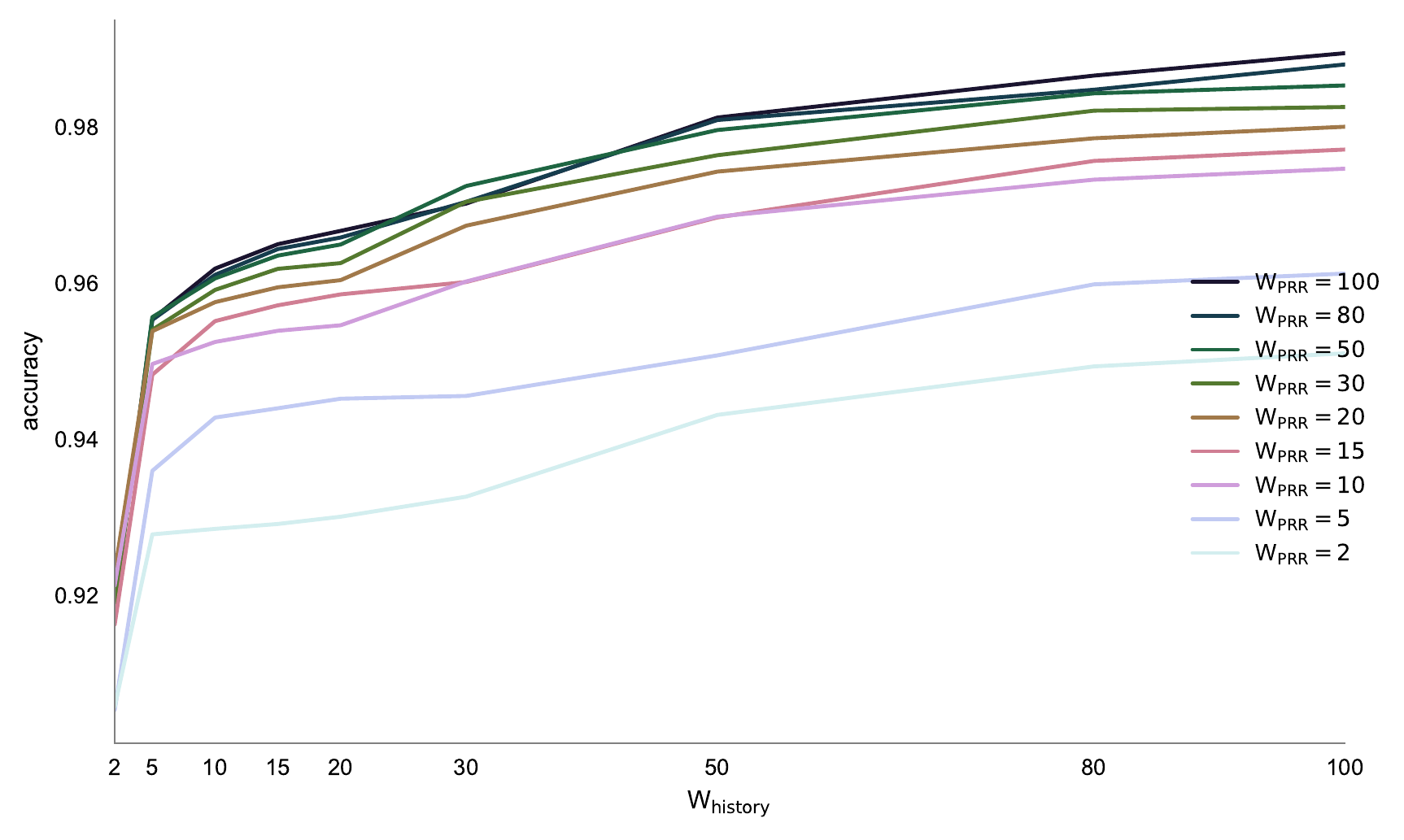}
	\caption{Overview of the influence of a discrete set of window sizes on the accuracy of the proposed nonlinear model.}
	\label{fig:window-sizes-linechart}
\end{figure*}

For examining the influence of the window selection on the performance of the model, we need to distinguish between two types of windows. The first one is the historical window $\Whistory$ that is used for computing features such as $\RSSIavg$. The second one is the prediction window $\Wprr$ that is used for computing the link quality labels. The majority of related works mention details about the window selection step. However, many of them fail to specify the size of the window used for the models they propose and evaluate. Additionally, the window size tends to be smaller for more reactive or online models, such as investigated in~\cite{liu2011foresee, rekik2015fli}, while for less reactive models, as proposed in~\cite{shu2017research, demetri2019automated}, the window size is likely larger.

Given that the investigated Rutgers trace-set consists of 300 packets per link, the size limits for the two windows are within $[0, 300]$ packets, where opting for the value 0 indicates no windowing and favoring the value 300 suggests per link labeling. Therefore, we restrict the range of the window sizes to $[2, 100]$ packets, within which we investigate the performance with a discrete set of nine values \{2, 5, 10, 15, 20, 30, 50, 80, 100\}. In this analysis, we predict the link quality for the next prediction window $\PRR(\Wprr)$ considering the Rutgers trace-set with domain knowledge interpolation, the  decision tree algorithm, with stratified k-fold and pruning, the feature vector ($\RSSI$, $\RSSIavg(\Whistory)$, $\RSSIstd(\Whistory)$), standard normalization and the random oversampling approach.

As portrayed in Fig.~\ref{fig:window-sizes-linechart}, the best performing model is the one utilizing $\Wprr=100$, which predominantly outperforms the models based on other $\Wprr$ settings, although all results for window size above 30 are rather similar. 

The results, in general, reveal that; (i) a longer historical window improves prediction because there is more information about how the link performed in the past, and (ii) increasing the prediction window (computing the future value of the classes for link quality) also leads to an improvement of the accuracy. Both observations, however, can also be a side-effect of ``smoothing''/averaging data from a relatively static trace-set. More explicitly, larger prediction windows are unable to inform on short-term effects, although they can help better in identifying the overall link behavior. It is worth noting that the optimal combination of values for historical and prediction windows is data dependent, however, the trade-offs discussed in this section can be adopted for general models. While the Rutgers trace-set is relatively static, for a more dynamic trace-set the optimal window sizes are likely smaller. 

To develop a suitable LQE model, the agility of the model has to be specified by the designer considering dynamically changing environments, e.g., for designing a routing algorithm in a largely mobile wireless network. Additionally, the practical memory limitations of the devices have to be taken into account when developing a suitable LQE model. This is mainly because more agile estimators use smaller window sizes, and therefore they tend to consume less memory, and yet yield low accuracy. Even though larger window sizes assist in attaining high accuracy, the cold start period, during which the historical window is initialized, leads to an estimation delay.

\subsection{Re-sampling strategy}
\label{sub:resampling}

\begin{figure}[!htb]
	\centering
	\subfloat[No resample]{\includegraphics[width=0.33\linewidth]{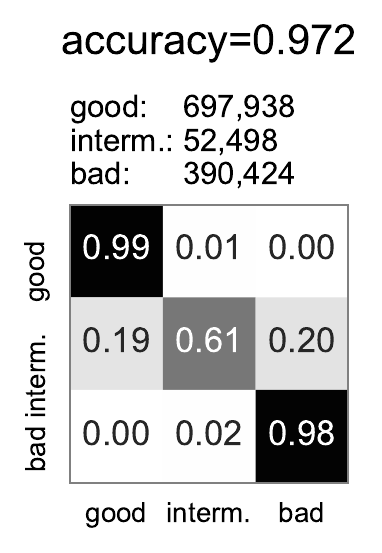}\label{fig:resample:none}}
	\subfloat[Undersampling]{\includegraphics[width=0.33\linewidth]{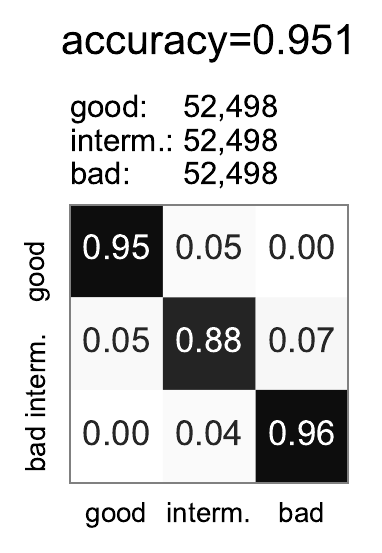}\label{fig:resample:under}}
	\subfloat[Oversampling]{\includegraphics[width=0.33\linewidth]{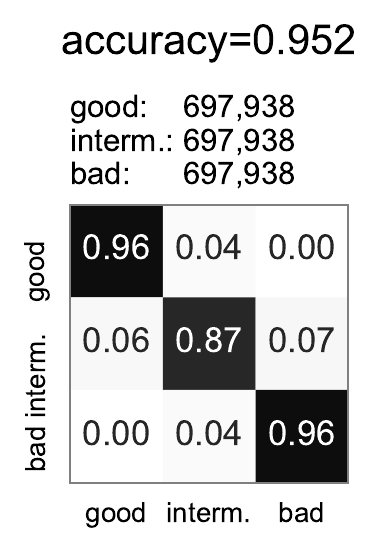}\label{fig:resample:over}}
	
	\caption{Different re-sampling strategies on the pipeline with a standard normalization and nonlinear decision tree algorithm using ($\RSSI$, $\RSSIavg$ and $\RSSIstd$ features.}
	\label{fig:resampling}
\end{figure}

From the analysis of the actual values in the considered Rutgers trace-set, it can be readily observed that there are 61\% \textit{good}, 34\% \textit{bad} and only 5\% \textit{intermediate} class entries. This distribution of data is largely imbalanced due to the artifact of the experiment, where the nodes were close to each other and the interference level was relatively low. Therefore, the majority of the links were actually good as expected and this was not due to the missing values within one particular class category of link quality. Additionally, it has been acknowledged in the literature~\cite{baccour2012radio} that the \textit{intermediate} region of the receivers tends to be relatively narrow compared to the \textit{good} and \textit{bad} regions, and therefore this naturally forms a scarcely populated class for intermediate regions in such trace-sets, yet having an important influence to ML-based LQE models although, as mentioned in the introduction as part of the motivation for this work, this aspect has been neglected by all the related work we have reviewed.

Imbalanced trace-sets are often encountered in ML and data mining communities and they are typically dealt with an appropriate re-sampling strategy. For studying the influence of the re-sampling strategy on the performance of the model for link quality classification, we employ the standard ROS and the RUS approaches. The ROS~\cite{chawla2004special, yap2014application} approach equalizes all class sizes to the size of the majority class by duplicating the trace-set entries of the minority classes; therefore the resulting re-sampled dataset becomes larger. On the contrary, the RUS~\cite{chawla2004special, yap2014application} approach equalizes all class sizes to the size of the minority class by randomly discarding instances from other larger classes. Hence, the new resampled dataset becomes smaller. It is observed from the numbers of good, intermediate and bad links in Figs.~\ref{fig:resampling}(b) and (c) that with both approaches, i.e., ROS and RUS, we are able to acquire a training dataset with balanced classes, about 50k examples for each class for ROS and 690k examples for ROS.

Fig.~\ref{fig:resampling} illustrates that re-sampling strategies on the Rutgers trace-set decrease the overall accuracy of the classification model from 97.2\% to slightly above 95\%. Some more advanced re-sampling strategies\cite{lusa2013smote} may limit this decrease in performance. However, when no re-sampling is performed, the minority class, i.e., \textit{intermediate} is only correctly detected in 61\% of the instances, indicating that the model is over-fitted to the majority of the classes. \textit{In the case of re-sampling, the minority class is correctly detected in over 87\% of the instances, yielding more than 25 percentage points increase in the performance.} This improvement comes at a relatively small performance cost for the majority classes, inducing 3-4 percentage points decline for the \textit{good} links and 2 percentage points reduction for the \textit{bad} links.

Considering this analysis, we may hint that, in the case of \cite{luo2019link}, where the performance of the predictor on two of the five classes is modest, employing a resample strategy might lead to better discrimination of those classes. Re-sampling may also improve other proposed estimators, for example the ones in~\cite{liu2012talent, liu2014temporal, shu2017research, demetri2019automated}.

The results for the selected Rutgers trace-set reveal that there is no significant distinction between the two re-sampling strategies, i.e., RUS and ROS. This is likely due to the relatively large size of the intermediate class. Although the intermediate class only represents 5\% of the population, it still contains more than 52,000 samples. However, looking beyond this particular trace-set, the RUS approach may suffer from excluding a certain number of majority class instances and may affect the representativeness of the remaining data points, especially for more dynamic trace-sets. On the other hand, due to the enlarged number of data points, the ROS approach requires more computing resources for building a model. Note that the results obtained in this section are based on interpolation and cleaning using domain knowledge, instant $\RSSI$, $\RSSIavg$ and $\RSSIstd$ as features and $\Wprr$ and $\Whistory$ of size 10. 

\section{Analysis of model selection}
\label{sub:building-model}
The final step of this systematic analysis is concerned with the influence of the ML algorithm selection on the performance of LQE models. To provide a comparative analysis of the impact, we examine logistic regression and linear SVM as representatives of linear ML algorithms, and decision trees, random forests and a multilayer perceptron, that is a class of feed-forward neural networks, as representatives of nonlinear model. As a baseline reference model, we leverage the majority classifier, which in our case, classifies all links in the \textit{good} class. 

\begin{figure}[!tbh]
	\centering
	
	\subfloat[Majority classifier]{\includegraphics[width=0.33\linewidth]{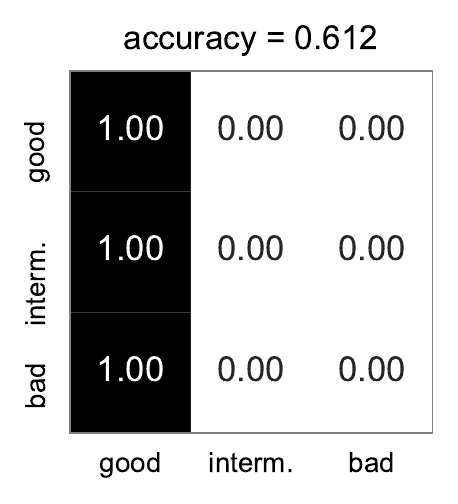}\label{fig:models:baseline}}%
	~
	\subfloat[Logistic regression]{\includegraphics[width=0.33\linewidth]{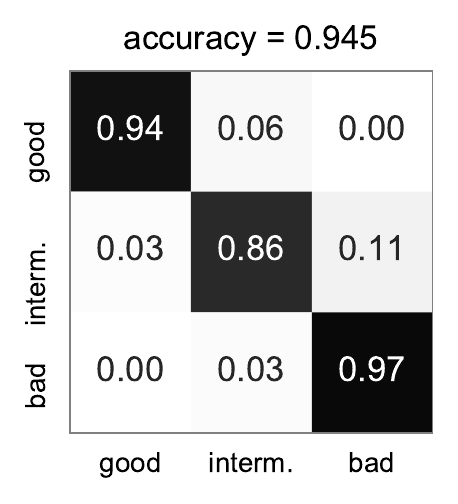}\label{fig:models:logreg}}%
	~
	\subfloat[Decision Trees]{\includegraphics[width=0.33\linewidth]{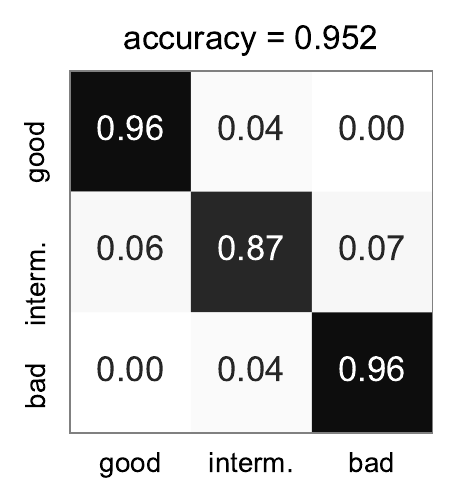}\label{fig:models:dtree}}%

	\subfloat[Random Forest]{\includegraphics[width=0.33\linewidth]{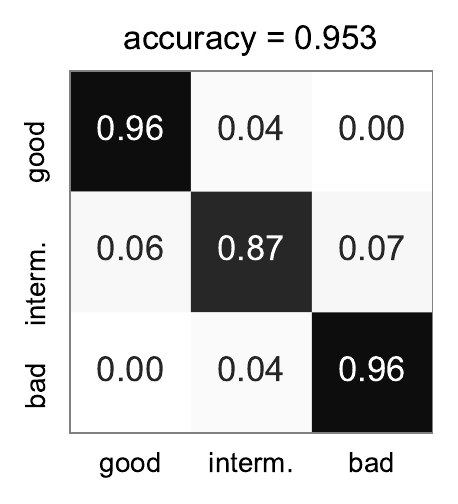}\label{fig:models:randforest}}%
	~
	\subfloat[SVM classifier]{\includegraphics[width=0.33\linewidth]{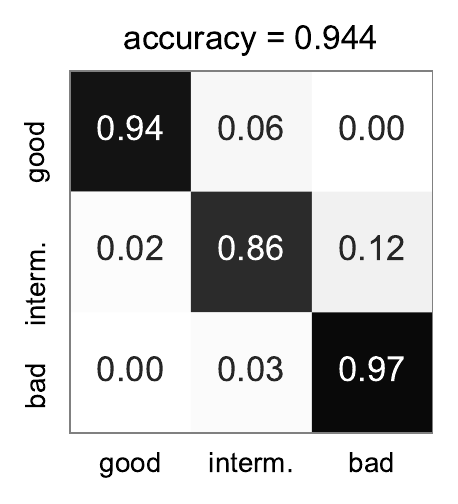}\label{fig:models:svm}}%
	~
	\subfloat[Multilayer perceptron]{\includegraphics[width=0.33\linewidth]{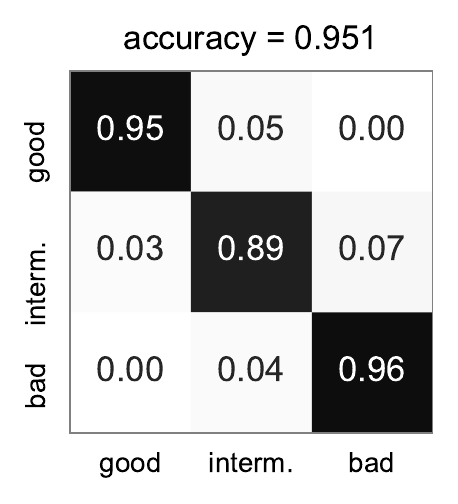}\label{fig:models:mlp}}%
	
	\caption{The influence of the choice of ML algorithm on the effectiveness of LQE models.}
	\label{fig:models}
\end{figure}

The analysis in this section is conducted by using domain knowledge interpolation, the feature vector consisting of instant $\RSSI$, $\RSSIavg$ and $\RSSIstd$, windowing with $\Wprr=10$, $\Whistory=10$, and a random oversampling approach over the Rutgers trace-set. The selected ML algorithms are evaluated using 10-times stratified K-fold cross-validation~\cite{arlot2010survey, pedregosa2011scikit}. Note that we obey the rule of cumulative parameterization throughout the data preprocessing steps in order to reveal the impact of each step on the ML algorithms for the sake of the LQE model proposed.

\textit{Fig.~\ref{fig:models} shows that all the selected ML models apart from the reference majority classifier have comparable performance, with an accuracy above 94\%.} Decision tress, random forests and multi-layer perceptrons, non-linear ML models, are very similar at 95\% accuracy. SVM with linear kernel and logistic regression are then at 94\%. Slightly lower performance of the linear models such as logistic regression and SVM conforms to the findings in the literature that LQE is a nonlinear function~\cite{baccour2012radio, millan2015time, bote2018online, shu2017research}. Looking at the ability of the ML algorithms to identify the minority class, the multilayer perceptron outperforms all the other ML algorithms considered for this analysis. 

One of our major observation from the analysis of ML-based LQE models is that the slightly better performance of nonlinear ML-based LQE models to the linear counterparts conforms to the findings in the state-of-the-art literature as it can be observed in~\cite{millan2015time, bote2018online}. Besides, upon the conclusions drawn in~\cite{liu2011foresee, liu2012talent, liu2014temporal, rekik2015fli}, which are mainly compared to 4B~\cite{fonseca2007four}, we can see that ML-based LQE models consistently outperform the traditional analytical estimators.

\section{Conclusions}
\label{sec:conclusion}
In this paper, we provided a systematic quantification of the influence of the design steps on the final performance of a wireless link quality classifier. Among others, we found that, for the chosen problem and dataset, the generation of synthetic features from the only available training feature $RSSI$, yields up to 6\% higher accuracy and is able to discriminate the intermediate class up to 49\% better. The choice of ML method has relatively smaller impact on final model performance with all the selected algorithm yielding accuracy between 94\% and 95\% and minority class is detected between 87\% and 89\%.

We also provided a first time evaluation of the impact of re-sampling on wireless link quality classification using ML. In the case of the chosen imbalanced dataset, by using standard re-sampling, the minority class was correctly detected in over 87\% of the instances, yielding more than 25 percentage points increase in the performance and comes at a small decrease in accuracy that can be mitigated with more advanced re-sampling techniques.

While some of the numbers will differ for other comparable datasets, the general conclusion is that balancing the dataset and carefully engineering the features provide more benefit than that of using very sophisticated and computationally expensive ML algorithms.

\section*{Acknowledgment}
This work was funded in part by the Slovenian Research Agency (Grant no. P2-0016 and J2-9232) and in part by the EC H2020 NRG-5 Project (Grant no. 762013).

\bibliographystyle{IEEEtran}
\bibliography{ftr_eng}

\end{document}